\documentclass[a4paper,12pt]{article}
\pdfoutput=1
\usepackage{feynmp-auto,expdlist}
\usepackage{amsmath, amsfonts, amssymb}
\usepackage{graphicx}
\usepackage{enumerate,cancel}
\usepackage{hyperref}
\usepackage{latexsym}
\usepackage{dsfont}
\usepackage{hepnicenames}
\usepackage{enumerate}
\usepackage{soul}
\usepackage[normalem]{ulem}
\usepackage{comment}

\usepackage{units}

\usepackage{soul}

\newcommand{\fNL}{f_{\rm NL}}

\usepackage{mathrsfs,graphicx,rotating,amsmath,amsfonts,mathtools,booktabs,amssymb,wasysym}
\usepackage{hyperref}\usepackage{slashed}
\usepackage[nosort]{cite}
\usepackage[table,xcdraw,s]{xcolor}
\usepackage{bm}
\usepackage{graphicx}
\usepackage{multirow,multicol}
\hypersetup{colorlinks,bookmarksopen,bookmarksnumbered,
	linkcolor=blus,pdfstartview=FitH,urlcolor=rossos,citecolor=verde}
\allowdisplaybreaks

\renewcommand{\[}{\left[}

\newcommand{\Lag}{\mathscr{L}}

\newcommand{\mio}[1]{}

\newcommand{\med}[1]{\langle #1\rangle}

\newcommand{\bpm}{\begin{pmatrix}}
\newcommand{\epm}{\end{pmatrix}}

\usepackage{mathrsfs}

\newcommand{\fig}[1]{~\ref{fig:#1}}

\allowdisplaybreaks
\usepackage{multicol}
\usepackage{color}
\definecolor{rosso}{cmyk}{0,1,1,0.4}
\definecolor{rossos}{cmyk}{0,1,1,0.55}
\definecolor{rossoc}{cmyk}{0,1,1,0.2}
\definecolor{blu}{cmyk}{1,1,0,0.3}
\definecolor{blus}{cmyk}{1,1,0,0.6}
\definecolor{bluc}{cmyk}{1,1,0,0.1}
\definecolor{verde}{cmyk}{0.92,0,0.59,0.25}
\definecolor{verdec}{cmyk}{0.92,0,0.59,0.15}
\definecolor{verdes}{cmyk}{0.92,0,0.59,0.4}

\newcommand{\bp}{\bar{M}_{\rm Pl}}
\renewcommand\&{&}

\newcommand{\eq}[1]{~{\rm (\ref{eq:#1})}}

\newcommand{\SR}{\cancel{\rm SR}}

\newcommand{\beq}{\begin{equation}}
\newcommand{\eeq}{\end{equation}}

\newcommand{\bea}{\begin{eqnarray}}
\newcommand{\eea}{\end{eqnarray}}
\newcommand{\be}{\begin{equation}}
\newcommand{\ee}{\end{equation}}
\font\tenrsfs=rsfs10 at 12pt
\font\sevenrsfs=rsfs7
\font\fiversfs=rsfs5
\newfam\rsfsfam
\textfont\rsfsfam=\tenrsfs
\scriptfont\rsfsfam=\sevenrsfs
\scriptscriptfont\rsfsfam=\fiversfs
\def\l{\left}
\def\r{\right}
\def\ddd{\mathrm{d}}
\def\be#1\ee{\begin{equation}#1\end{equation}}
\def\bl#1\el{\begin{align}#1\end{align}}
\def\ba#1\ea{\begin{align*}#1\end{align*}}

\renewenvironment{thebibliography}[1]
{\begin{multicols}{2}[\section*{\refname}]%
		\@mkboth{\MakeUppercase\refname}{\MakeUppercase\refname}%
		\list{\@biblabel{\@arabic\c@enumiv}}%
		{\settowidth\labelwidth{\@biblabel{#1}}%
			\leftmargin\labelwidth
			\advance\leftmargin\labelsep
			\@openbib@code
			\usecounter{enumiv}%
			\let\p@enumiv\@empty
			\renewcommand\theenumiv{\@arabic\c@enumiv}}%
		\sloppy
		\clubpenalty4000
		\@clubpenalty \clubpenalty
		\widowpenalty4000%
		\sfcode`\.\@m}
	{\def\@noitemerr
		{\@latex@warning{Empty `thebibliography' environment}}%
		\endlist\end{multicols}}

\newcommand{\eV}{\,{\rm eV}}

\makeatletter
\font\ital=cmu10

\def\hhref#1{\href{http://arxiv.org/abs/#1}{arXiv:#1}}
\usepackage{xstring}
\newcommand{\hhrefq}[1]{\IfSubStr{#1}{:}{\href{http://inspirehep.net/search?ln=en&ln=en&p=#1&of=hb&action_search=Search&sf=&so=d&rm=&rg=25&sc=0}{InSpire:#1}}{\hhref{#1}}}

\def\art{\@ifnextchar[{\eart}{\oart}}
\def\eart[#1]#2#3#4#5#6{{\rm #2}, {\em #3 \bf #4} {\rm (#6) #5} ({\em #1})}
\def\article{\@ifnextchar[{\earticle}{\oarticle}}
\def\oarticle#1#2#3#4#5#6{{\rm #1}, {\ital `#6'}, {\rm #2 #3 (#5) #4}}
\def\earticle[#1]#2#3#4#5#6#7{{\rm #2}, {\ital `#7'}, {\rm #3 #4 (#6) #5}  [\hhrefq{#1}]}
\def\hepart[#1]#2{{\rm #2, \sl#1}}
\def\heparticle[#1]#2#3{#2, {\ital `#3'} [\hhrefq{#1}]}
\newcommand{\doi}[1]{\href{http://dx.doi.org/#1}{[link]}}
\def\oarticle#1#2#3#4#5#6{{\rm #1}, {\rm #2 #3 (#5) #4}}
\def\earticle[#1]#2#3#4#5#6#7{{\rm #2}, {\rm #3 #4 (#6) #5}  [\hhrefq{#1}]}
\def\heparticle[#1]#2#3{#2 [\hhrefq{#1}]}

\newcommand{\hhrefqq}[1]{\IfBeginWith{#1}{10.}{\href{https://doi.org/#1}{doi:#1}}{\hhrefq{#1}}}

\renewenvironment{thebibliography}[1]
{\begin{multicols}{2}[\section*{\refname}]%
		\@mkboth{\MakeUppercase\refname}{\MakeUppercase\refname}%
		\list{\@biblabel{\@arabic\c@enumiv}}%
		{\settowidth\labelwidth{\@biblabel{#1}}%
			\leftmargin\labelwidth
			\advance\leftmargin\labelsep
			\@openbib@code
			\usecounter{enumiv}%
			\let\p@enumiv\@empty
			\renewcommand\theenumiv{\@arabic\c@enumiv}}%
		\sloppy
		\clubpenalty4000
		\@clubpenalty \clubpenalty
		\widowpenalty4000%
		\sfcode`\.\@m}
	{\renewcommand{\@noitemerr}
		{\@latex@warning{Empty `thebibliography' environment}}%
		\endlist\end{multicols}}

\newcounter{alphaequation}[equation]
\renewcommand{\thealphaequation}{\theequation\hbox to
	0.6em{\hfil\alph{alphaequation}\hfil}}
\definecolor{Gray}{gray}{0.95}

\usepackage{tocloft}
\begin{document}
\thispagestyle{empty}
\begin{center}  
{\LARGE\bf\color{rossos} Traversing a kinetic pole during inflation:\\
primordial black holes and gravitational waves} \\
\vspace{0.6cm}
{\bf Anish Ghoshal}$^a$ and {\bf Alessandro Strumia}$^b$  \\[6mm]
{\it $^a$
Institute of Theoretical Physics, Faculty of Physics, University of Warsaw, Poland}\\
{\it $^b$ Dipartimento di Fisica, Universit\`a di Pisa, Pisa, Italia}\\[1mm]

\vspace{0.5cm}
{\large\bf Abstract}
\begin{quote}\large
We consider an inflationary kinetic function with an integrable pole
that is traversed during inflation.
This scenario leads to enhanced spectra of primordial scalar inhomogeneities with detectable signals:
formation of primordial black holes
(that could explain Dark Matter) and
scalar-induced gravitational waves (that could
reproduce the recent Pulsar Timing Array observation, or predict signals in future detectors such as LISA or ET).
Spectral signatures depend on whether the inflaton mass dimension 
at the pole is above or below 2.
Values mildly below 2 allow a big power spectrum enhancement with a mild tuning.
Finally, we discuss the possibility that a kinetic pole can arise as anomalous
dimension of the inflaton due to quantum effects of Planckian particles that become
light at some specific inflaton field value.
\end{quote}
\end{center}
\setcounter{page}{1}
\tableofcontents

\newpage

\section{Introduction}


Models of inflation have been obtained under the
assumption that the kinetic term of an inflaton scalar $\phi$ contains a {\em non-integrable} 
pole at some field value
$\phi= \phi_{\rm pole}$
\beq \Lag = K(\phi) \frac{(\partial_\mu \phi)^2}{2}- V(\phi) ,\qquad K (\phi) \sim \left|\frac{\phi_*}{\phi - \phi_{\rm pole}}\right|^p,\qquad p \ge 2.\eeq
Indeed, after canonicalisation of $\phi$ into $\phi_{\rm can}$, the Lagrangian simplifies to
 \beq \Lag =  \frac{(\partial_\mu \phi_{\rm can})^2}{2}- V_{\rm can}(\phi_{\rm can})\eeq
where the scalar potential   $V_{\rm can}$ undergoes an {\em infinite} stretch,
acquiring the flatness needed to support inflation. 
Pole inflation models have been motivated invoking
string theory, supergravity~\cite{1405.3646,1506.00936,1505.03386,1703.00305}
or a non-minimal scalar coupling $f(\phi) R$ to gravity~\cite{2109.10367}.

\medskip

We here consider a milder {\em integrable} pole with $0<p<2$,
equivalent to a {\em finite} stretch of the inflaton potential,
and thereby to an {\em exact} inflection point in the canonical potential.
This can give interesting signals if $\phi$ is the inflaton.
Indeed it is known that an {\em approximate} inflection point in an inflationary potential,
characterized by a small $V'$,  
leads to a phase with an amplified power spectrum of primordial inhomogeneities 
$P_\zeta \propto 1/V'^2$ which subsequently gives rise to
scalar-induced gravitational waves and to primordial black holes~\cite{Ivanov:1994pa,gr-qc/0503017,0801.0116,1510.05669,1610.09362,1705.06225,1706.04226,1706.05007,1709.05565,1712.09750,1805.06731,1805.09483,1810.12608,1811.10108,1905.13581,1910.10602,2001.05909,2111.01362,2207.11878,2306.04002,2305.12325}.\footnote{Inflection-point inflation~\cite{1601.05979,2104.03977} and hybrid inflation~\cite{2402.06613} are also also possible.}
We will discuss that an {exact} inflection point, as obtained from a traversable integrable pole,
also gives these signals.

In section~\ref{can} we present the correspondence between the kinetic function $K(\phi)$, the potential $V(\phi)$ 
and the canonical potential $V_{\rm can}(\phi_{\rm can})$.
In section~\ref{sec:formalism} we set up the formalism to conveniently compute keeping $K\neq 1$,
avoiding the canonicalisation that trades a pole $K(\phi_{\rm pole})=\infty$ for an inflection point in $V_{\rm can}$.
In section~\ref{sec:power} we compute the predicted power spectrum $P_\zeta(k)$.
In section~\ref{GW} we compute the resulting frequency spectrum $\Omega_{\rm GW}(f)$
of gravitational waves generated
at second order in $P_\zeta$.
In section~\ref{BH} we determine the resulting mass spectrum of primordial black holes. 
Section~\ref{0<1<p} will show that $p<1$ allows to obtain such signals without a large tuning of model parameters.

\smallskip

In section~\ref{theory} we introduce a potential theoretical explanation for the existence of an integrable pole.
Some particles with Planck-scale mass (such as string states) might become massless 
at a point $\phi= \phi_{\rm pole}$ in field space, during the inflaton's Planckian journey in field space.
Light particles induce quantum corrections to the kinetic function of $\phi$ that can lead to a pole-like structure,
with a value of $p$ related to the anomalous dimension of $\phi$.
Some particle interactions contribute as $p<0$, leading to a vanishing $K$ rather than to a pole:
after canonicalisation this is equivalent to a jump-like feature in the inflaton potential:
we also mention the consequent signals~\cite{jump,2104.03972}.

Conclusions are given in section~\ref{end}.

%
%

 \section{Potential for the canonical inflaton}\label{can}
We work in the Einstein frame, where scalars have
minimal coupling to gravity (so that the graviton kinetic term is canonical),
and can have a non-canonical kinetic function $K(\phi)$ in the action:
\beq \label{eq:genericlagrangian}
S= \int d^4 x \sqrt{|\det g|} 
\bigg[  - \frac{\bar M_{\rm Pl}^2}{2} R+ \frac{K(\phi)}{2} {(D_\mu \phi)(D^\mu \phi)}-V(\phi) + \cdots\bigg].\eeq
We assume one scalar $\phi$ with one pole at $\phi=\phi_{\rm pole}$.
With one scalar only, the non-canonical  kinetic term $K$  can be reabsorbed by defining a canonically normalised scalar $\phi_{\rm can} (\phi)$ as $ d\phi_{\rm can}/{d\phi } = \sqrt{K}$.
In this section we discuss how the pole is equivalent to a feature in the canonical potential 
\beq V_{\rm can}(\phi_{\rm can})=
V(\phi(\phi_{\rm can})).\eeq
We assume a kinetic function with a pole ($p>0$) or a dip ($p<0$):
\beq \label{eq:Kpole} 
K(\phi) =\left\{ \begin{array}{ll}
\displaystyle 1+ \left| \frac{\phi_*}{\phi - \phi_{\rm pole}}\right|^{p} & \hbox{for }p> 0\\
\displaystyle
\left[ 1+ \left| \frac{\phi_*}{ \phi -\phi_{\rm pole}}\right|^{-p}\right]^{-1} & \hbox{for } p < 0
\end{array}\right.
\stackrel{\phi\to \phi_{\rm pole}}\simeq \left|\frac{\phi_*}{\phi - \phi_{\rm pole}}\right|^{p}.
\eeq
The term $1$
makes the kinetic term canonical, away from the special point $\phi_{\rm pole}$.\footnote{A related study~\cite{2001.05909,2111.01362} considered a different form of $K$ with $p=1,6/5$ motivated by Brans-Dicke gravity.}
The full $\phi_{\rm can}(\phi)$ resulting from the $K(\phi)$ in eq.\eq{Kpole}
can be written in terms of hypergeometric functions.
We will later more simply compute inflation without performing the canonicalisation,
and we here discuss the effect of canonicalisation around the 
special point $\phi_{\rm pole}$, where one can neglect this constant term in $K$, obtaining for any $p$
\beq
\label{eq:canoninfl}
\frac{\phi_{\rm can}-\phi_{\rm pole}}{\phi_*} \simeq d	 \left|\frac{\phi  - \phi_{\rm pole}}{\phi_*} \right| ^{1/d} ,\qquad
d \equiv \frac{2}{2-p}=\left\{\begin{array}{ll}
1& \hbox{for }p=0\\
2 & \hbox{for }p=1\\
3 &\hbox{for }p=4/3
\end{array}
\right.  ~.
\eeq
Eq.\eq{canoninfl} means that
the parameter $d$ is the dimension of the field $\phi$ around the pole.
In section~\ref{theory} we will
interpret $d\neq 1$ as a quantum anomalous dimension.

\subsection{Kinetic function with a pole, $p>0$}
Let us first consider the pole case, corresponding to $p>0$.
In line with the literature on pole inflation, we make the assumption that the pole in the kinetic term 
is not accompanied by a non-trivial feature in the potential. As a result, the potential can be approximated through a first-order Taylor expansion. Therefore, we proceed with the assumption that, around the pole,
\begin{align}
\label{eq:poleaction}
V(\phi)\simeq V_0 + V'_0(\phi-\phi_{\rm pole}) .
\end{align}
Significant changes occur at two critical values: $p=1$ and $p=2$.

A pole with $p\ge 2$ is non-integrable: 
the $\phi$ region near the pole becomes an infinite range of the canonical $\phi_{\rm can}$ field,
so the potential gets infinitely stretched acquiring a nearly-flat inflationary form~\cite{1405.3646,1506.00936,1505.03386,1703.00305,2109.10367
}.\footnote{A non-integrable pole can arise, in the Jordan frame, from a canonical kinetic term and
a non-minimal coupling to gravity $f(\phi) R$
with a function $f$ that crosses zero at $\phi_{\rm pole}$.
The two regions $\phi<\phi_{\rm pole}$ and $\phi>\phi_{\rm pole}$ are disjoint,
so that the problematic side where $f<0$ does not affect the side where $f>0$~\cite{2109.10367}.}

\begin{figure}[t]
\begin{center}
$$\includegraphics[width=0.9\textwidth]{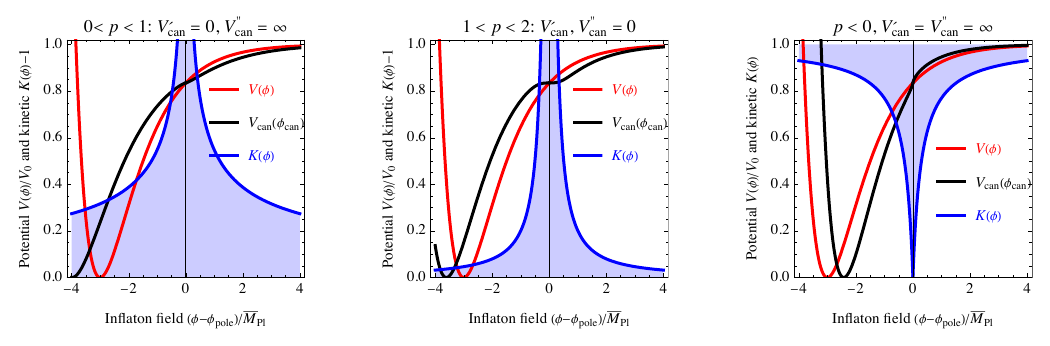}$$
\caption{\em\label{fig:SampleVSemiPole} 
How a kinetic function $K(\phi)$ as in eq.\eq{Kpole} (in blue) modifies an
inflaton potential $V(\phi)$ as in eq.\eq{Vinfl}
(in red), rewriting the system as a potential $V_{\rm can}(\phi_{\rm can})$ (in black)
for the canonical field $\phi_{\rm can}$.
{\bfseries Left}: traversable pole with low $p<1$, such that
$V_{\rm can}'=0$ and $V_{\rm can}''= \infty$ at the pole.
{\bfseries Middle}: traversable pole with $p=4/3$ such that
$V_{\rm can}$ is cubic around the inflection point.
{\bfseries Right}: dip with $p<0$ such that $V_{\rm can}$ develops a jump.
The position $\phi_{\rm pole}$ of the pole and its thickness $\phi_*$ are free parameters.}
\end{center}
\end{figure}

This paper focuses instead on an integrable pole with $0<p<2$, that gives a finite
stretching of the potential around the pole, where the potential
acquires the form of an exact inflection  point 
\beq \label{eq:Vinflection}
V_{\rm can}(\phi_{\rm can}) \simeq
V_{\rm pole}(\phi_{\rm can}) \equiv  
V_0 \left[1+ \left(\frac{\phi_{\rm can}-\phi_{\rm pole}}{\phi_0}\right)^d\right] \qquad\hbox{where}\qquad
\phi_0= \phi_* d \left(\frac{V_0}{V'_0\phi_*}\right)^{1/d}
. \eeq
We introduced the parameter $\phi_0$ and
left understood how to deal with non-integer $d$ at $\phi_{\rm can}<\phi_{\rm can}^{\rm pole}$.
The first derivative vanishes at the pole, 
$V'_{\rm can}(\phi_{\rm pole})=0$, so the associated slow-roll parameter $\epsilon_V \propto V'^2_{\rm can}$ vanishes.
In the slow-roll approximation the power spectrum $P_\zeta  = H^2/8\pi^2 \bp^2\epsilon_V$ diverges,
hinting that the slow-roll approximation breaks down and $P_\zeta$ is enhanced.
Two sub-cases need to be considered:
\begin{itemize}
\item If $1<d<2$ (corresponding to $0<p<1$) 
the  number of $e$-folds computed in slow-roll approximation (eq.\eq{efolds} later)
remains finite, and the second derivative 
$\eta_V\propto V''_{\rm can}$ diverges at $\phi_{\rm pole}$
breaking the slow-roll approximation.
The modified kinetic term stretches the field space also away from $\phi_{\rm pole}$.
The left panel of fig.\fig{SampleVSemiPole} shows an example.

\item If $d> 2$ (corresponding to $p>1$) the pole has a local effect,
producing an exact inflection point where  $V'_{\rm can}= V''_{\rm can}=0$
and thereby both slow-roll parameters vanish.
The middle panel of fig.\fig{SampleVSemiPole} shows an example with $d=3$.
The pole is only traversed beyond the slow-roll approximation, that breaks down.
The $V_{\rm can}\simeq V_{\rm pole}$ approximation 
only holds in a field region with 
size $\Delta \phi \sim \phi_*$ around the pole.
Outside this region $K\simeq 1$ so that
$V_{\rm can} (\phi_{\rm can}) \simeq V(\phi + \Delta)$ where
$\Delta$ is some finite shift due to the stretching occurred around the pole.

\end{itemize}
Assuming that $V(\phi)$ globally has inflationary form, 
we can now discuss the inflationary consequences of a traversable pole.
If  the $V_{\rm pole}$ approximation of eq.\eq{Vinflection} were applicable away from the pole,
the slow-roll parameters would get large at $|\phi_{\rm can}-\phi_{\rm pole}| \gtrsim \phi_0$.
However it's important to note that eq.\eq{Vinflection} only holds in a narrow field range with size
$\Delta \phi$ around the pole.
We will be interested in $\Delta \phi \ll \phi_0$, such that the slow-roll parameters can be always small.

\subsection{Kinetic function with a dip, $p<0$}
The kinetic function $K$ vanishes at the special point $\phi_{\rm pole}$ if $p<0$.
Infinitesimally near to $\phi_{\rm pole}$ the canonical potential is approximated 
by eq.\eq{Vinflection} where now $0<d<1$.
This means that the canonical potential develops a
jump, where $V'_{\rm can}=V''_{\rm can}=\infty$. 
An example of such a scenario is displayed in the right panel of fig.\fig{SampleVSemiPole}.

We anticipate that a jump
tends to produce a mild oscillatory feature in  the power spectrum $P_\zeta(k)$.
However, in certain cases, a more significant amplification of $P_\zeta$ can occurr:
a) from a series of sharp jumps~\cite{jump,2104.03972},
which would correspond to a series of poles with $p\ll -1$.
b) from a very small value of the slow-roll parameter $\epsilon$ away from the pole,
as possible in small-field inflation~\cite{2104.03972}.
This would give distinctive features in $P_\zeta(k)$, that get imprinted in the gravitational wave spectrum. 
For the remainder of this paper, we will focus on cases where $p>0$
and avoid pursuing these dip possibilities.

\section{The formalism for computing inflation}\label{sec:formalism}
In this section we recall the standard formul\ae{} that will be later
used to compute inflation in the theory of eq.\eq{genericlagrangian},
with a generic inflaton potential $V$ and a generic kinetic function $K$.
The $\Lambda$CDM best-fit values of the scalar tilt $n_s$, 
of the power spectrum of scalar perturbations $P_\zeta(k)$, and
of the tensor-to-scalar ratio $r=P_h/P_\zeta$ from {\sc Planck}-BICEP 
at the CMB pivot scale $k\approx 0.056/{\rm Mpc}$
are~\cite{CMBdata}
\beq \label{eq:infpredex}
n_s = 0.9649 \pm 0.0042 ,\qquad P_\zeta \approx (2.1 \pm 0.06) \times 10^{-9},\qquad
r\lessapprox 0.035\eeq
around cosmological scales corresponding to $N\approx 50-60$ $e$-folds before the end of inflation.
Larger ranges are found in $\Lambda$CDM extensions.
The classical inflaton equation of motion is
\beq \label{eq:clast}
\ddot\phi + \frac{K'}{K} \frac{\dot\phi^2}{2}+3H \dot\phi + \frac{V'}{K}=0 .
\eeq

\subsection{The slow-roll approximation}
The slow-roll approximation neglects 
the first two terms in eq.\eq{clast}.
In the slow-roll limit, the slow-roll parameters are given by~\cite{Burns:2016ric,Karamitsos:2017elm}
\beq
\label{eq:srp}
\epsilon_V \equiv -\frac{\dot H}{H^2}\simeq
\frac{1}{2} \frac{\bar{M}_{\rm Pl}^2}{ K} \frac{V'^2}{V^2} ,\qquad
\eta_V \equiv \frac{\dot\epsilon_V}{H\epsilon_V}\simeq
\frac{\bar{M}_{\rm Pl}^2}{K} \frac{V'}{V} \frac{d\ln\epsilon_V}{d\phi}.
\eeq
Inflation ends when $\epsilon_V(\phi_{\rm end}) \approx 1$ and the number
of $e$-foldings is given by
\begin{align}
\label{eq:efolds}
N (\phi) = \int_{\phi_{\rm end}}^{\phi }d\phi' \  \frac{K(\phi')}{\bar{M}_{\rm Pl}^2} \frac{V(\phi')}{V'(\phi')}.
\end{align}
The predictions for $n_s$, $P_\zeta$ and $r$ are well known, 
\begin{align}\label{eq:infpredth}
n_s = 1 -2 \epsilon_V+\eta_V,\qquad
P_\zeta = \frac{1}{24\pi^2} \frac{V}{\bar{M}_{\rm Pl}^4 \epsilon_V},\qquad
r = 16\epsilon_V
\end{align}
evaluated around horizon crossing.
The power spectrum of eq.\eq{infpredth} diverges at $\phi_{\rm pole}$
as the kinetic function diverges, $K\to \infty$, leading to $\epsilon_V\to 0$.
Furthermore, this point is non-traversable in slow-roll approximation if $p>1$, as eq.\eq{efolds} diverges.
In any case the slow-roll approximation
breaks down around the pole when $\eta_V\sim 1$ corresponding to $(V'/\sqrt{K})'/\sqrt{K}\sim H^2$.

\subsection{Classical motion beyond the slow-roll limit}
We thereby solve the full classical equation of motion,
using the number of $e$-folds $dN=H\, dt$ rather than time $t$ as variable:
\beq \label{eq:classeom}
\frac{1}{3-\epsilon_H}\left(K\frac{d^2\phi}{dN^2} + \epsilon_H  \frac{\bp^2 K'}{K}\right)
+K\frac{d\phi}{dN} 
+\bp^2 \frac{V'}{V}=0\eeq
where
\beq \label{eq:Heps}
H^2=\frac{V}{\bp^2(3-\epsilon_H)},\qquad
\epsilon_H \equiv-
\frac{d\ln H}{dN}=
\frac{K}{2\bp^2}\left(\frac{d\phi}{dN}\right)^2\qquad\hbox{and}\qquad
\eta_H\equiv \epsilon_H - \frac{1}{2\epsilon_H} \frac{d\epsilon_H}{dN}
\eeq
will be used later. 
Eq.\eq{classeom} will be numerically solved starting from the slow-roll initial condition $d\phi/dN=-\bp^2V'/KV$.
This provides the approximation for $n_s$, $P_\zeta$ and $r$ known as `exact slow-roll'. 
Following the notations of~\cite{1905.13581}:
\beq  \label{eq:PzetaH}
n_s=1-4\epsilon_H+2\eta_H,\qquad
P_\zeta = \frac{1}{24\pi^2} \frac{V}{\bar{M}_{\rm Pl}^4 \epsilon_H},\qquad
r=16 \epsilon_H
.\eeq
This power spectrum qualitatively differs from the slow-roll approximation in eq.\eq{infpredth} at the pole point where $\epsilon_V=0$ while $\epsilon_H\neq 0$.

\subsection{Mukhanov-Sasaki equations}
To precisely compute the power spectrum $P_\zeta(k)$ one needs to expand the field $\phi$ 
into fluctuating modes $\phi_k$
with co-moving momentum $k$ and solve the equations for their inflationary evolution.
As usual it is convenient to consider the gauge-invariant combination 
$\zeta = -\psi+H \delta \phi/\dot\phi$
of the inflaton $\phi$ and the gravity potential $\psi$.
The power spectrum is obtained as $P_\zeta = k^3|\zeta_k|^2/2\pi^2$ from
the values of the complex modes $\zeta_k$ at inflation end.
Defining $v_k = - z \zeta_k$ where
$z^2 = 2 a^2 \epsilon_H$ and $a=e^N$ is the scale factor,
the Mukhanov-Sasaki (MS) equation for $v_k$ is\footnote{The MS equation was derived in~\cite{hep-th/0605045} in terms of conformal time $\tau$ in a theory with generic inflaton action $P(X,\phi)$ where $X=g^{\mu\nu}(\partial_\mu\phi)(\partial_\nu\phi)/2$ is the kinetic term:
\beq \label{eq:MSgen}
\frac{d^2 v_k}{d\tau^2} + \left(c_s^2 k^2 - \frac{1}{z}\frac{d^2 z}{d\tau^2}\right) v_k=0,\qquad
z = \frac{a\sqrt{2\epsilon}}{c_s}\qquad \epsilon_H = \frac{P_{,X} X}{M_{\rm Pl}^2 H^2},\qquad
P_\zeta = \frac{H^2}{8\pi^2 \bp^2 c_s\epsilon_H}.
\eeq
Our model corresponds to $P=K(\phi) X - V(\phi)$ giving $c_s=1$, so that  $\epsilon_H$ and $P_\zeta$ in eq.\eq{MSgen}
reduce to eq.\eq{Heps}.}
\beq\label{eq:MS}
\frac{d^2 v_k}{dN^2}+(1-\epsilon_H) \frac{dv_k}{dN} + \left[\frac{k^2}{a^2 H^2} +(1+\epsilon_H-\eta_H)(\eta_H-2)-\frac{d(\epsilon_H-\eta_H)}{dN}\right] v_k = 0.\eeq
Eq.\eq{MS} can be numerically solved starting from the Bunch-Davies initial condition
\beq v_k = \frac{1}{\sqrt{2k}},\qquad \frac{dv_k}{dN}=\frac{k v_k}{iaH}.\eeq
In numerical computations it is more convenient to solve the equation for $\zeta_k$
\beq \frac{d^2 \zeta_k}{dN^2}+ f \frac{d\zeta_k}{dN} + 
\frac{k^2}{a^2 H^2} \zeta_k=0,\qquad
f = 3+ \epsilon_H- 2\eta_H
\eeq
because no $\phi'''$ is involved.
 A negative $f<0$ signals amplification of $\zeta_k$.
The number of $e$-folds $N =\ln  k/k_{\rm end}$ is converted
into the wave-number $k$ by computing $k_{\rm end}$,
the scale corresponding to inflation end,
assuming instantaneous reheating with temperature
$T_{\rm RH}=(30 V(\phi_{\rm end})/\pi^2 g)^{1/4}$ where $g=106.75$ is the number of Standard Model
degrees of freedom.
Standard big-bang cosmology then gives the scale factor at inflation end,
$a_{\rm end}=a_{\rm eq} (g_{s\rm eq}T_{\rm eq}/g T_{\rm RH})^{1/3}$
in terms of $g_{s\rm eq}=43/11$, $T_{\rm eq}\approx 0.74\eV$, $a_{\rm eq}\approx 1/3100$.

\subsection{The inflationary potential}
As a concrete example, we assume the inflationary potential
\beq \label{eq:Vinfl}
V = V_0 \left[1-e^{-\sqrt{2/3}\phi/\bp}\right]^2\eeq
independently motivated by $R^2$ gravity~\cite{Starobinsky:1980te},
 by $\alpha$-attractor models~\cite{1405.3646,1506.00936,1505.03386,1703.00305}, 
and by Higgs-like scalar inflation with large non-minimal scalar coupling to gravity $\xi\gg 1$~\cite{0710.3755}.
Without a kinetic pole, the inflaton potential assumed in eq.\eq{Vinfl} leads to
$n_s=1-2/N$ and $r=12/N^2$, in agreement with $\Lambda$CDM best fits for $N\sim (50-60)$.


\section{The power spectrum}\label{sec:power}
We now use the formalism of section~\ref{sec:formalism} to compute 
inflation in the theory of eq.\eq{genericlagrangian}.
Following the classification of section~\ref{can}
a pole with $1<p<2$ is considered in section~\ref{1<p<2},
while $0<1<p$ is considered in section~\ref{0<1<p}.

\begin{figure}[t]
\begin{center}
$$\includegraphics[width=\textwidth]{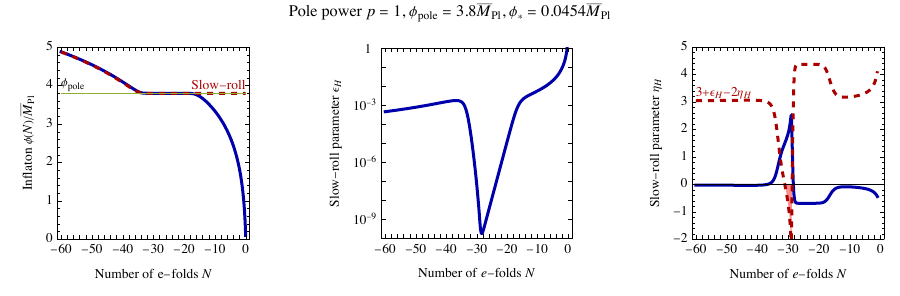}$$
\caption{\em\label{fig:Inflation} 
Sample of inflationary evolution as function of the number of $e$-folds $N$ for pole power $p=1$.
Left: the inflaton field $\phi$ compared to the slow-roll approximation.
Middle: the $\epsilon_H$ parameter.
Right: the $\eta_H$ parameter and $f=3+\epsilon_H-2\eta_H$.
Amplification of $P_\zeta$ occurs for $f<0$, in the shaded region.
} 
\end{center}
\end{figure}

\subsection{Pole exponent $1\le p<2$}\label{1<p<2}
The number of $e$-folds in slow-roll approximation 
$N \propto \int d\phi ~K$ diverges around a pole with $p\ge 1$,
corresponding to an inflection point, as exemplified by the dashed curve in the left panel of  fig.\fig{Inflation}.
So a pole with $1\le p<2$ is not traversed in slow-roll approximation,
and we need to solve the full eq.\eq{classeom}.

\smallskip

We first present a simple analytic argument showing that the pole is traversed
as long as the pole width is sub-Planckian, $\phi_* \lesssim 0.1\bp$.
Making the field canonical the pole is transformed into an inflection
point that can be roughly approximated as a field range 
\beq \label{eq:Deltaphi}
\Delta\phi_{\rm can}=\int_{-\phi_{\rm max}}^{+\phi_{\rm max}} d\phi\, (\sqrt{K}-1) \sim
\phi_* \left[\frac{4}{2-p}+\frac{1-(\phi_*/\phi_{\rm max})^{p-1}}{p-1}\right]\qquad
\hbox{for $0<p<2$}
\eeq
within which
the potential is constant, $V'_{\rm can}=0$ and $K_{\rm can}=1$.
The slow-roll approximation breaks down when the inflaton enters the flat region
with initial speed $\dot\phi_{\SR} \simeq \sqrt{2\epsilon_{\SR}} H \bp$
where $\epsilon_{\SR} \ll 1$ is the $\epsilon_V$ slow-roll parameter of the
inflationary potential $V$ just away from the pole.
Inside the nearly-flat pole region, the inflaton proceeds inertially
according to $\ddot\phi_{\rm can}+3H \dot\phi_{\rm can}=0$,
evolving by an amount $\phi_{\rm walk}=\dot\phi_{\SR}/3H$
before being stopped by Hubble friction.
Thereby the flat interval is traversed if 
\beq\label{eq:trav}
 \Delta\phi_{\rm can} \lesssim \phi_{\rm walk} \approx \sqrt{2\epsilon_{\SR}}\bp/3.\eeq
This means that a sub-Planckian pole $\phi_* \ll \bp$  is traversed.
After entering the flat region at $N=N_{\SR}$ $e$-folds,
the inflaton slows down increasing 
the power spectrum as $P_\zeta \sim P_{\zeta\SR} \,\exp (6(N - N_{\SR}))$.
A large enhancement of $P_\zeta$ arises
if the inflaton significantly slows down, corresponding to eq.\eq{trav}
being nearly saturated.
Above this value the classical approximation breaks down as $P_\zeta \sim 1$.
We are interested in $P_\zeta$ enhanced up to $10^{-2}$ 
as needed to form an allowed amount of primordial black holes
and/or generating a detectable allowed amount of gravitational waves.
We thereby select the value of $\phi_*$ in eq.\eq{Kpole}
such that  $P_\zeta(k) \lesssim 10^{-2}$.
Table~\ref{tab:FT} shows that such a large enhancement in $P_\zeta(k)$
needs a tuning of $\phi_*$ of order $(P_\zeta/10^{-9})^{1/2} \sim 10^3$~\cite{2304.01997}.

The shaded regions are excluded by the current
constraint on CMB $\mu$-distortion, $ |\mu| < 0.9~10^{-4}$~\cite{astro-ph/9605054} and by the effect on the neutron/proton abundance during big-bang nucleosynthesis~\cite{1605.04646}. The dashed curve
represents the future PIXIE sensitivity on $\mu$-distortion~\cite{1105.2044}, down to $ | \mu | \sim  10^{-7}$.

\medskip

\begin{figure}[t]
\begin{center}
$$\includegraphics[width=0.83\textwidth]{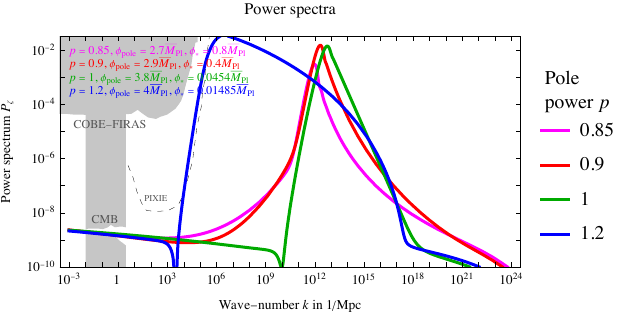}$$
\caption{\em\label{fig:SamplP} 
Sample of power spectra for different pole powers $p$.
For $p\ge 1$ the value of $\phi_*$ is tuned such that $P_\zeta$ reaches $\sim 10^{-2}$.
The gray regions are excluded by CMB, LSS, COBE-$\mu$, COBE-$y$, AR~\cite{1203.2681};
the dashed curve is the sensitivity of PIXIE~\cite{1105.2044}.
} 
\end{center}
\end{figure}

\begin{table}[t]
$$\begin{array}{c|ccc}
\multicolumn{4}{c}{\hbox{Fine tuning $\Delta_x = \partial \ln  P_\zeta^{\rm peak}/\partial \ln x$}}\cr \hline
\hbox{Sample} &\multicolumn{3}{c}{\hbox{Theory parameters}}\cr
\hbox{spectrum} &x= \phi_*  & x=\phi_{\rm pole} & x=p \cr \hline
\color{red}p=0.9 &\Delta=4 &\Delta=15&\Delta=35\cr
\color{verdes}p=1 &\Delta=25&\Delta=80 &\Delta=165\cr
\color{blue}p=1.2 &\Delta=130&\Delta=400&\Delta=400 \cr
\end{array}
$$
\caption{\label{tab:FT}\em Fine-tuning sensitivities $\Delta_x$ of $P_\zeta^{\rm peak}= \max_k P_\zeta(k)$
to theory parameters $x$ for the sample spectra that achieve $P_\zeta^{\rm peak}\approx 0.01$
with different values of the pole power $p$.}
\end{table}

Fig.\fig{SamplP} shows examples of the power spectra for a pole with the critical power
$p=1$  (corresponding to $d=2$, a quadratic canonical potential around the pole)
and for other values of $p$ around 1.
Fig.\fig{Inflation} shows additional details of the inflationary evolution in the $p=1$ example.
The power spectrum, obtained solving the MS equation,
is well approximation by eq.\eq{PzetaH} which, in turn,
differs from the approximation of eq.\eq{infpredth} only for the
narrow spike at the pole.
This $p=1$ example gives $n_s \approx 0.952$ and $r\approx 0.003$ at the CMB pivot scale,
as can be understood by noticing that the inflaton pausing around the pole
reduces its field value corresponding to the CMB scale.
These values are compatible with simple $\Lambda$CDM extensions (e.g.~\cite{2302.03268}), 
and the best-fit $\Lambda$CMD value of $n_s$ could be obtained by
assuming a different potential above the integrable kinetic pole.
A suitable choice is a generic potential made inflationary by an extra non-integrable pole with $p'>2$,
see eq.~(10) of~\cite{2109.10367}.
Table~\ref{tab:FT} shows that the $p=1$ critical case is less strongly tuned than $p>1$.

\smallskip

The behaviour of $P_\zeta(k)$ is understood as follows:
\begin{enumerate}
\item While $\phi$ approaches $\phi_{\rm pole}$,
$P_\zeta (k)$ quickly grows roughly proportionally to $e^{6N} \propto k^6$.

\item Around the pole $P_\zeta(k)$ reaches a maximal value well approximated by eq.\eq{PzetaH},
$\max P_\zeta \approx V_0/24\pi^2 \bp^2 \epsilon_H^{\rm min}$.
The minimal value of $\epsilon_H$  depends on the $\phi_*$ parameter.
Furthermore, the scale $k$ at which $P_\zeta$ is maximal depends on the value of $\phi_{\rm pole}$.

\item After crossing the pole, 
$P_\zeta$ slowly decreases as the  inflaton reaccelerates.
A new slow-roll approximation shows that this phase lasts
$\Delta N \sim(\phi_0/\bp)^{2/p} (\epsilon_H^{\rm min})^{(1-p)/p}$  $e$-folds.
An enhancement of $P_\zeta $ up to  $10^{-2}$ needs a small $\epsilon_H^{\rm min}$.
So a too long $\Delta N\gtrsim50$ is avoided if $p$ is mildly close to 1:
$p\lesssim 4/3$ corresponding to a cubic  $d=3$ inflection point in $V_{\rm can}$.
The numerical result in fig.\fig{SamplP} confirms 
that, for $p=1.2$ and for the inflationary potential of eq.\eq{Vinfl}, 
$P_\zeta$ can peak at small enough $k\lesssim1/{\rm kpc}$,
while satisfying COBE-FIRAS bounds~\cite{1203.2681} and giving signals in PIXIE~\cite{1105.2044}.

\end{enumerate} 
For the values of $1<p<2$ considered here the kinetic function $K$ deviates from 1 
only in a narrow region around the pole, so that inflationary predictions away from the pole are not affected by the pole.


\subsection{Pole exponent $0<1<p$}\label{0<1<p}
If $p<1$ the kinetic function $K$ deviates from 1 also away from the pole,
as clear from the fact that the field range of eq.\eq{Deltaphi} receives a contribution from $\phi_{\rm max}$.
As a result $P_\zeta$ and $n_s$ at CMB scales are affected by the pole,
and the traversability approximation of eq.\eq{trav} no longer applies.
Rather, the pole is always traversed, even in slow-roll approximation.
Increasing the parameter $\phi_*$ no longer allows to arbitrarily increase $P_\zeta$
at the price of a tuning.

Numerical solutions show that a significant enhancement
of $P_\zeta$ still arises for $p$ mildly below 1, $0.85\lesssim p<1$.
Fig.\fig{SamplP} shows examples for $p=0.9$ and for $p=0.85$.
The value $p=0.85$ only leads to a mild enhancement in $P_\zeta$, 
while a large enhancement up to $0.01$ still arises for $p =0.9$.
Table~\ref{tab:FT} shows an interesting feature of the $p=0.9$ example:
the logarithmic sensitivities with respect to all  parameters 
of the theory are mild, indicating the large enhancement
of $P_\zeta$ is achieved without a strong tuning.
This is interesting, given that several single-field inflation models lead to 
a strong enhancement of the primordial power spectrum at small scales
at the price of a strong fine-tuning \cite{2304.01997},
related to the fact that the enhancement
typically requires an ultra-flat region of the scalar field potential over a small field range.\footnote{Extra fields present during inflation may help in reducing the fine-tuning, see e.g.~\cite{2305.12325,2310.04174}.}
The tuning issue also affects other models, like those with transient inflationary features in the primordial power spectrum or new phase transitions (see~\cite{2301.03600} for a review).

%
%

\smallskip

We also see that $p<1$ gives peaks in $P_\zeta(k)$ with a different shape than $p>1$:
a slower growth before the peak, a shaper peak, a faster decrease after the peak.
The critical case $p=1$ results in a shape that is intermediate 
between the characteristic shapes associated with $p>1$ and $p<1$.
This shape will have an impact on the spectrum of gravitational waves, which will be the subject of discussion in the next section.

\section{Scalar-induced gravitational waves}\label{GW}
The primordial spectrum of scalar inhomogeneities $P_\zeta(k)$ induces
a spectrum of gravitational waves with density
$\Omega_{\rm GW}(f) \sim 10^{-5} P_\zeta^2(2\pi f)$~\cite{gr-qc/0612013,hep-th/0703290,1804.08577,1801.05235, 2109.01398, Escriva:2022duf}.
We start by summarising in section~\ref{sec:GWsummary} how  $\Omega_{\rm GW}$ is precisely computed,
and present results in section~\ref{resGW}.

\subsection{The formalism for computing gravitational waves}\label{sec:GWsummary}
First-order inflaton cosmological perturbations, described by the power spectrum $P_\zeta$  of 
curvature scalar perturbations,
induce gravitational waves, often referred to as Scalar Induced Gravitational Waves (SIGW),
at second order in cosmological perturbation theory \cite{Ananda:2006af, hep-th/0703290}.  
These gravitational waves are described by transverse-traceless tensor perturbations $h_{ij}$ in the metric.
This gravitational effect is conveniently computed
using the Newtonian gauge where the inflaton is homogeneous and the first-order scalar perturbations are described by one potential $\psi$, while neglecting anisotropic stress.
The metric is 
 \be\label{inf:metric_pert}
\ddd s^2 = -(1+2\psi)\ddd t^2 + a(t)^2 \l[(1-2\psi)\delta_{ij} + h_{ij} \r] \ddd x^i \ddd x^j ~,
\ee
The $h_{ij}$ of eq.~\eqref{inf:metric_pert} are inevitably induced via non-linear coupling between different modes of the first-order scalar modes of $\psi$.
At second perturbative order, the tensor modes with helicity  $\lambda$ evolve in conformal time $\tau$ as 
\be \label{eom_hk}
h_{\bm{k}}^{\lambda\prime\prime}(\tau) + 2aH h_{\bm{k}}^{\lambda\prime}(\tau) + k^2h_{\bm{k}}^{\lambda}(\tau) = S^\lambda_{\bm{k}}(\tau) ~,
\ee
where  the source term $S^\lambda_{\bm{k}}$ during radiation domination 
is
\beq \label{eq:source}
	S^\lambda_{\bm{k}}	=
	2 \int {\ddd^3\bm{p} \over (2\pi)^{3/2}} \bm{e}^\lambda(\bm{k},\bm{p})
	\Big[
	3 \psi_{\bm{p}} \psi_{\bm{k} - \bm{p}}
	+ \frac{\psi_{\bm{p}}' \psi'_{\bm{k} - \bm{p}}}{a^2{H}^{2} }
	+ \frac{\psi_{\bm{p}}' \psi_{\bm{k} - \bm{p}}
	+  \psi_{\bm{p}} \psi'_{\bm{k} - \bm{p}}}{a{H} }
	\Big] ~,
\eeq
$\bm{e}^\lambda(\bm{k},\bm{p}) \equiv e^\lambda_{lm}(\bm{k}) p_l p_m$
and all $\psi$ are evaluated at $\tau$.
Substituting $\psi_k =2 \zeta_k/3$~\cite{hep-ph/9807278},  solving eq.~\eqref{eom_hk} 
via the Green function method and summing over polarizations gives the SIGW power spectrum~\cite{1804.08577, Espinosa:2018eve} 
\be\mathcal{P}_h(\tau,k)
= \int^\infty_0 \mathrm{d}v \int^{|1+v|}_{|1-v|}\mathrm{d}u \l[ {4v^2-(1+v^2-u^2)^2 \over 4uv} \r]^2 I^2(v,u,k\tau) \mathcal{P}_\zeta(ku) \mathcal{P}_\zeta(kv) 
\ee
where $u \equiv |\bm{k} - \bm{p}|/k$, $v \equiv p/k$.
The kernel function $I^2$ evaluated at late time $\tau$ and averaged over time
(denoted with an overline) is~\cite{1804.08577, Espinosa:2018eve}
\be
\begin{aligned}
	\overline{I^2(v,u,k\tau \rightarrow \infty)} =& 
	 \left( \frac{3(u^2+v^2-3)}{ u^3 v^3 k\tau}\right)^2 \bigg[ \bigg(-4uv+(u^2+v^2-3) \ln\left| \frac{3-(u+v)^2}{3-(u-v)^2}\right| \bigg)^2  
	\\&
	+ \pi^2(u^2+v^2-3)^2 \Theta(v+u-\sqrt{3})\bigg] .
\end{aligned}
\ee
The dominant contribution to SIGWs is produced at $\tau_k$ just after the re-entry of modes inside the Hubble horizon,
as scalar perturbations damp away quickly when sub-horizon during radiation domination~\cite{Espinosa:2018eve}.  
The GW energy density 
$\rho_{\text{GW}}(\tau,\bm{x}) =  \bp^2  \langle h'_{ij}(\tau,\bm{x}) h^{ij}{}'(\tau,\bm{x}) \rangle/4 a^2(\tau)$~\cite{Brill:1964zz, Isaacson:1968hbi, Isaacson:1968zza, Ford:1977dj, Ota:2021fdv} 
is usually presented as
\be \label{omega_gw}
\Omega_{\rm GW} (\tau, k)\equiv
\frac{1}{\rho_{\rm tot}}\frac{d\rho_{\rm GW}}{d\ln k}
= \frac{1}{12} \l( {k \over a H} \r)^2 \overline{\mathcal{P}_h(\tau,k)} ~.
\ee
Eq.~\eqref{omega_gw} applies while radiation dominates, $\Omega_{\rm rad}(\tau_k)\simeq 1$, and the thermal bath consists of  $g(\tau_k)$ degrees of freedom.
As the universe cools, some degrees of freedom may decouple, and after the matter-radiation equality epoch, radiation experiences a more significant redshift compared to matter. This ultimately leads to the present-day density of SIGW,
\begin{equation}
\frac{\Omega_{\rm GW}(\tau_0)}{\Omega_{\rm rad}(\tau_0)} = \frac{a^4(\tau_k)\rho_{\rm rad}(\tau_k)  }{a^4(\tau_0)\rho_{\rm rad}(\tau_0)} \frac{\Omega_{\rm GW}(\tau_k) }{\Omega_{\rm rad}(\tau_k)}
=\left(  \frac{g_{s}(\tau_0)}{g_{s}(\tau_k)}\right)^{4/3}  
\frac{g(\tau_k)}{g(\tau)}  \Omega_{\rm GW}(\tau_k)\,.
\end{equation}
Inserting $\Omega_{\rm rad}(\tau_0) h^2 = 4.1 ~ 10^{-5}$ and assuming $g(\tau_k)=g_{s}(\tau_k)$ gives
\be
\Omega_{\rm GW}(\tau_0, f) h^2 \approx 1.6 ~ 10^{-5} 
 \l( {106.75 \over g(\tau_k)  } \r)^{1/3}
\Omega_{\rm GW}(\tau_k, k),
\ee
in terms of the present-time frequency $f = k/2\pi $ as $a(\tau_0)=1$.

\begin{figure}[t]
\begin{center}
$$\includegraphics[width=0.85\textwidth]{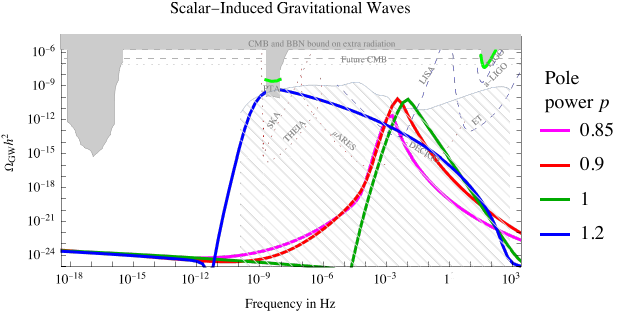}$$
\caption{\em\label{fig:PlotGWPoleInflation} 
Frequency spectra of
scalar-induced gravitational waves corresponding to the $P_\zeta(k)$ of fig.\fig{SamplP}
for the indicated pole powers $p$.
The regions shaded in gray are excluded.
The hatched regions are below the expected astrophysical foregrounds, that could be partially subtracted.
The dashed (dotted) curves show the sensitivities of planned (futuristic) experiments.
The green curves are the detections from LIGO/VIRGO and Pulsar Timing Arrays.
} 
\end{center}
\end{figure}

\subsection{Gravitational waves: results}\label{resGW}
Fig.\fig{PlotGWPoleInflation} displays sample gravitational wave spectra  $\Omega_{\rm GW}(\tau_0, f)$
corresponding to the sample curvature power spectra shown in fig.\fig{SamplP}.
The increase of $P_\zeta$ with $k$ (proportional to $k^6$ for $p>1$) is fast enough that SIGW
can reproduce the nHz gravitational wave signal claimed by Pulsar Timing Arrays~\cite{PTAs}.
This is illustrated by the $p=1.2$ curve in fig.\fig{PlotGWPoleInflation}.
For lower $p$ we choose values of $\phi_{\rm pole}$ such that the
SIGW spectrum peaks at higher frequencies, possibly in the range to be explored
by LISA or ET. 

Fig.\fig{PlotGWPoleInflation} also includes shaded gray regions denoting the current SIGW exclusion bounds:
\begin{itemize} 
\item The CMB constraint on $r$ of eq.\eq{infpredex} implies the
strong bound on $\Omega_{\rm GW}$ at low $f\sim 10^{-16}\,{\rm Hz}$; 

\item The BBN and CMB bound on the energy density $\rho_{\rm GW}$ of extra radiation applies at any $f$
and is usually presented in terms of an effective number of extra neutrinos~\cite{Maggiore:1999vm}
\beq \Delta N_{\rm eff}=4.4\left.\frac{\rho_{\rm GW}}{\rho_\gamma}\right|_0 =1.8 ~ 10^5
\int_{f_\text{min}}^{\infty} \frac{\text{d}f}{f}   \Omega_\text{GW}(f) h^2 
\approx  10^5     \Omega_\text{GW}^\text{peak} h^2 . \label{eq:darkrad2} \eeq
The current bound $\Delta N_{\rm eff}\lesssim  0.28$~\cite{Planck:2018vyg}
could be improved by 1 or 2 orders of magnitude with future data~\cite{2203.05728}.
The lower integration limit is $f_\text{min}\approx 10^{-10}\,\text{Hz}$ for BBN 
and $f_\text{min}\approx 10^{-18}\,\text{Hz}$ for CMB. 
Ignoring the precise frequency dependence on the spectrum, this sets 
the bound $\Omega_{\rm GW}^{\rm peak}\lesssim 10^{-5}$ plotted in fig.\fig{PlotGWPoleInflation}
on the peak value of $\Omega_{\rm GW}(f)$.

\item LIGO-VIRGO-{\sc KaGra} interferometer observations at $f\sim {\rm kHz}$~\cite{LIGOs}; 

\item Pulsar Timing Array (PTA) observations at $f\sim {\rm nHz}$~\cite{PTAs}.
\end{itemize}
The green curves in fig.\fig{PlotGWPoleInflation} show the gravitational wave signals claimed by
LIGO-VIRGO-{\sc KaGra} 
and by PTA.
These signals are likely attributed, for the most part, to the mergers of astrophysical black holes.
The PTA signal could also be due to a stochastic background of gravitational waves,
as PTA do not (yet) observe individual sources or events.

Fig.\fig{PlotGWPoleInflation} also shows the expected sensitivities of some planned future experiments:
advanced LIGO and VIRGO~\cite{VIRGO:2014yos,LIGOScientific:2019lzm} 
and the Einstein Telescope (ET) \cite{Hild:2010id} at $f\gtrsim {\rm Hz}$;
the space-based interferometer
\textsc{LISA}~\cite{Baker:2019nia} at $f\sim{\rm mHz}$.
Additionally, the figure includes some futuristic experiments:
DECIGO \cite{Kawamura:2011zz}, $\mu$-ARES \cite{Sesana:2019vho} at $f\sim {\rm mHz-Hz}$,
and
the SKA PTA~\cite{Weltman:2018zrl}
and the \textsc{THEIA} star survey~\cite{Garcia-Bellido:2021zgu} at $f\sim{\rm nHz}$.
Even ignoring more futuristic experiments, improvements by many orders of magnitude  in the sensitivity to gravitational waves seem possible.

However, astrophysical gravitational wave foregrounds act as backgrounds that might limit the 
sensitivity to detect primordial gravitational waves.
Fig.\fig{PlotGWPoleInflation} also shows the sum of the diffuse astrophysical backgrounds.
These estimates come with an uncertainty of about an order of magnitude and,
in the range observed by LIGO-VIRGO-{\sc KaGra},
can primarily be categorised
as either binary black hole (BH-BH) \cite{BHBH} or binary neutron star (NS-NS) \cite{LIGOs,Abbott:2017xzg} events.
Around Hz frequencies it might be possible to subtract such foregrounds,
 reaching sensitivities $\Omega_{\rm GW} \sim 10^{-13-15}$ \cite{Cutler:2005qq,Regimbau:2016ike}. 
At lower frequencies in the LISA range, 
the dominant binary white dwarf (WD-WD) foreground \cite{Farmer:2003pa, Rosado:2011kv, Moore:2014lga}  could be subtracted reaching $\Omega_{\rm GW} \sim 10^{-13}$~\cite{Adams:2010vc, Adams:2013qma}. 

In the optimistic scenario where foregrounds can be effectively subtracted to reach the anticipated sensitivities,
SIGW can be distinguished from the astrophysical foreground,
that exhibits an expected spectral shape ($f^{2/3}$ in some range~ \cite{Zhu:2012xw})
which is distinct from the spectral characteristics of SIGW.

\section{Primordial Black Holes}\label{BH}
In the preceding sections we showed how a traversable kinetic pole can enhance
the small-scale curvature perturbations,
resulting in regions with over-dense
$\delta \equiv \delta\rho/\rho$.
When a mode $k$ re-enters the horizon 
during the post-inflationary radiation-dominated epoch,
these over-dense regions may gravitationally collapse into Primordial Black Holes
(PBHs) if the density perturbation $\delta$ exceeds the threshold $\delta_{\rm th}$ of the collapse, 
estimated as $\delta_{\rm th}\approx 0.5$
in Press-Schechter approximation~\cite{Press:1973iz,gr-qc/0412063,Musco:2020jjb}.
In such a case a PBH forms with mass given by $\gamma$ times the mass inside the horizon~\cite{1801.05235}.
The order unity factor $\gamma\approx 0.2$ accounts for the estimated
efficiency of the gravitational collapse in spherical approximation~\cite{Carr:1975qj}. 
So the PBH mass $M_{\rm PBH}$ is~\cite{astro-ph/9901293}
\begin{equation}
   M_{\rm PBH}  \approx
   M_\odot  \frac{\gamma}{0.2}\l( \frac{2}{k\,{\rm pc}} \r)^{2}
   \approx \gamma \frac{ M_{\rm Pl}^2 } {H_{\rm infl}} e^{2 |N|} 
\end{equation}
in terms of $k$  or 
in terms of the number $N$ of $e$-folds before the end of inflation when
the mode $k$ leaves the horizon during inflation.

\subsection{The formalism for computing Primordial Black Holes}
The fraction of the universe volume collapsing into PBHs can be computed in terms of
the variance of the density perturbation $\delta$,
smoothed over a radius $R$ to control small-scale divergences~\cite{1801.05235}:
\begin{equation}
   \sigma^2(R)\equiv  \langle \delta^2 \rangle  = \int \mathrm{ d } \ln k  \, P_\delta (k) \,{W}^2 ( kR ) 
      = \frac{16}{81} \int \mathrm{ d } \ln k \, ( kR )^4 P_\zeta (k) {W}^2 ( kR )   .
\end{equation}
We choose a Gaussian smoothing $W(kR)=e^{-(kR)^2/2}$,
so that a given $R$ corresponds to $k \sim 1/R$,
and thereby
the size $R$ is related to the PBH mass $M_{\rm PBH}$ as $R \approx 2 G M_{\rm PBH}/ \gamma a \sim 1/aH$.
The fraction of the Universe ending up in PBHs is given by the $\delta>\delta_{\rm th}$ tail of the distribution for $\delta$,
approximated as Gaussian: 
\beq
\beta ( M_{\rm PBH} ) = \gamma \int^\infty_{\delta_{\rm th}} \mathrm{d} \delta\, P(\delta)
\stackrel{\sigma \ll \delta_{\rm th}}{\simeq}
        \frac{\gamma \sigma}{  \sqrt{2 \pi}  \delta_{\rm th} }\exp \left(  - \frac{\delta_{\rm th}^2} { 2\sigma^2} \right),     
\qquad     
      P(\delta) =  \frac{\exp\l( -\delta^2/2 \sigma^2 \r)}{\sqrt{2 \pi \sigma^2} }.
   \label{eq:betaM}
\eeq 
The total PBH density at formation is $\rho_\text{PBH}= \rho_{\rm tot} \int \beta(M) \ddd\ln M$.
The fraction $f_{\rm PBH} $ 
of PBH relative to the DM abundance at given $M_{\rm PBH}$ mass at current time is~\cite{1801.05235}
\begin{equation}
   f_{\rm PBH} \approx  2.7 ~ 10^{8} 
         \bigg( \frac{ 10.75}{ g_{*,\rm form}} \bigg)^{1/4} 
         \bigg(\frac{\gamma}{0.2} \frac{ M_\odot}{M_{\rm PBH}}  \bigg)^{1/2} 
         \beta   
\end{equation}
assuming adiabatic expansion of the cosmological background  after PBH formation and ignoring accretion.
The PBH abundance roughly scales as $e^{-1/P_\zeta}$ and 
is peaked at the value of $k$ that maximises $P_\zeta(k)$.
A PBH abundance comparable to the DM abundance arises
for $\max P_\zeta \sim 10^{-2}$, and
is exponentially sensitive to deviations from this value.

\smallskip

\begin{figure}[t]
\begin{center}
$$\includegraphics[width=0.31\textwidth]{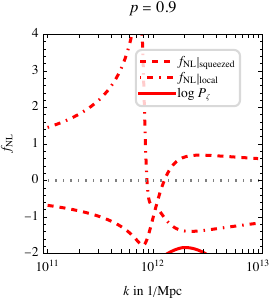}\quad
\includegraphics[width=0.31\textwidth]{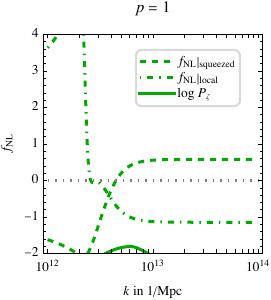}\quad
\includegraphics[width=0.31
\textwidth]{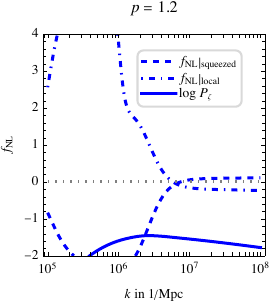}
$$
\caption{\em\label{fig:fNL} Bispectrum parameter $\fNL$ computed in the squeezed limit and in the local limit using the $\delta N$ approximation
around the peak of power spectrum $P_\zeta(k)$.
} 
\end{center}
\end{figure}

Since PBHs only form in regions with large over-density, their formation rate could be significantly affected by possible non-Gaussian tails in the distribution function of the density perturbations~\cite{1801.09415,1811.07857,2211.01728}.
However this remains an open problem, and the PBH abundance anyhow has a strong exponential dependence on model parameters.
In our current analysis we do not attempt including effects due to non-Gaussianity, viewing them as uncertainties.
Below we nevertheless discuss some impact of non-Gaussianities.
The first correction beyond the Gaussian approximation is encoded in the bispectrum $B_\zeta$ defined as~\cite{Byrnes:2010ft,Ade:2015ava} 
\begin{equation}\label{Bi}
\left\langle\zeta_{\bm{k}_{1}}\zeta_{\bm{k}_{2}}\zeta_{\bm{k}_{3}}\right\rangle=(2 \pi)^{3} \delta^{3}\left(\bm{k}_{1}+\bm{k}_{2}+\bm{k}_{3}\right) B_{\zeta}\left(k_{1}, k_{2}, k_{3}\right),
\end{equation}
For detailed expressions of bi-spectrum $B_\zeta(k_1,k_2,k_3)$ see~\cite{1201.0926,Arroja:2011yj,Zhang:2020uek}.
This is usually parameterised in terms of a dimension-less $\fNL$ parameter that depends on the modulus of three momenta,
typically defined as \cite{Creminelli:2006rz,Byrnes:2010ft}
\begin{equation}\label{Fnl}
f_{\text{NL}}(k_1,k_2,k_3)=\frac{5}{6}\frac{B_{\zeta}(k_1,k_2,k_3)}{P_{\zeta}(k_1)
P_{\zeta}(k_2)+P_{\zeta}(k_2)P_{\zeta}(k_3)+P_{\zeta}(k_3)P_{\zeta}(k_1)}.
\end{equation}
One-field models imply the general result   $\fNL|_{\rm squeezed} = 5 (1-n_s)/12$  in the squeezed limit $k_3\to 0$ and thereby $k_1=k_2$~\cite{astro-ph/0210603,astro-ph/0407059}.
In our examples the spectral index $n_s(k)$ of eq.\eq{PzetaH} crosses $1$ near the peak, and deviates from 1 around the peak of $P_\zeta$,
see fig.\fig{fNL}.
According to~\cite{2111.01362} this might suppress non-Gaussianities.
The $\delta N$ formalism dictates $\fNL$ in the local limit as
$f_{\rm NL}|_{\rm local}= - 5 \phi''/6\phi'$~\cite{astro-ph/0504045},
that too crosses 0 around the peak, see fig.\fig{fNL}.
A generic result for $\fNL(k_1,k_2,k_3)$ is provided in~\cite{1201.0926}.
The equilateral limit $k_1=k_2=k_3$ maybe relevant for PBH formation given that $P_\zeta(k)$ exhibits a narrow peak,
such that the PBH abundance gets approximatively multiplied by~\cite{2106.10792}
\beq \exp\left[ \frac{23\delta_{\rm th}^3}{P_\zeta(k_{\rm peak})} \fNL(k_{\rm peak},k_{\rm peak},k_{\rm peak})\right]. \eeq
According to~\cite{1801.09415}, around an inflection point 
the cubic momentum of the $P(\delta)$ distribution reduces its variance 
$\sigma^2$ by 1.6.
Overall, these considerations suggest $\fNL\sim 1$, so that leading-order non-Gaussianities play a significant role.

So higher order non-Gaussianities can be relevant too.
 A tentative all-order result writes $\zeta$ in terms of a Gaussian fluctuation $\zeta_{\rm G}$
 in the local limit in ultra-slow roll models as~\cite{1905.13202, 2211.01728}
 \beq \zeta =- \frac{\ln (1- 6 \fNL\zeta_{\rm G}/5)}{6\fNL/5} \simeq \zeta_{\rm G} + \frac35 \fNL \zeta_{\rm G}^2 + \cdots\eeq 
According to~\cite{2211.13932} ultra-slow-roll inflation can produce an exponential (rather than Gaussian) tail.

Non-Gaussianities also affect, in a milder way, the scalar-induced GW discussed in section~\ref{GW}.
A $\fNL \sim 1$ tends to give an order unity correction to their frequency spectrum, see e.g.~\cite{SIGWfNL}. 
However a fully non-Gaussian estimate is still a topic of active research, with new contributions being identified~\cite{2308.07155,SIGWfNL,2403.06962,2305.19950}. 
A dedicated computation of non-gaussianities is beyond the scope of the current analysis.



\begin{figure}[t]
\begin{center}
$$\includegraphics[width=0.85\textwidth]{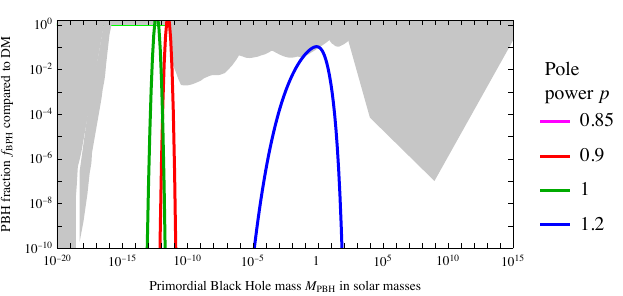}$$
\caption{\em\label{fig:PlotPBHPoleInflation} 
Mass spectra of primordial black holes corresponding to the $P_\zeta(k)$ of fig\fig{SamplP}
for the indicated pole powers $p$. The case $p=0.85$ predicts a 
low $f_{\rm PBH}$, not visible in the figure.
The normalization is significantly uncertain.
The areas shaded in grey are excluded, from~\cite{ReviewDM}.
} 
\end{center}
\end{figure}

\subsection{Primordial Black Holes: results}
Fig.\fig{PlotPBHPoleInflation} shows the PBH mass distributions produced by the sample power
spectra of fig\fig{SamplP}.
The case with $p=0.85$ gives a negligible PBH abundance that does not appear in the plot.
For larger $p$
we could choose parameter values such that $P_\zeta$ reaches $0.01$,
saturating the current bounds on the PBH abundance,
as indicated by the gray areas in fig.\fig{PlotPBHPoleInflation}.

For $p=0.9$ and $p=1$ we fixed the position of $\phi_{\rm pole}$ in such a way that the generated 
PBHs fall in the
mass range around $10^{-16-12}$ solar masses where PBHs can be all Dark Matter, $f_{\rm PBH}\approx 1$,
and the related SIGW are in the range to be explored by LISA.

On the other hand, larger $p$ only allows to generate heavier PBHs.
This is because  the inflaton remains a larger number $\Delta N$ of $e$-folds around $\phi_{\rm pole}$
after  having produced a peak $P_\zeta \approx 0.01$.
The sample power spectrum for $p=1.2$ generates SIGW around the PTA claim
and solar-mass PBH with density around current bounds. 
PBH with masses below a solar mass could be detected by LIGO/VIRGO, 
as they would not have astrophysical backgrounds, but these PBHs cannot account for all of the Dark Matter.

Non-Gaussianities are expected to affect both the PBH abundance and
the Scalar-Induced Gravitational Waves~\cite{SIGWfNL}.
A precise inclusion of non-Gaussianities would be interesting because:
i) reproducing the PTA observation as SIGW is expected to imply a significant amount of PBH~\cite{2306.17149};
ii) DM as PBH is expected to imply detectable SIGW.
We leave a detailed investigation of the impact on non-Gaussianities  for future work.

\section{Light particles during inflation}\label{theory}
To conclude, we present a possible theory motivation for
a kinetic function $K(\phi)$ with a traversable pole or with a dip, as in eq.\eq{Kpole}.

\smallskip

A noteworthy aspect of large-field inflation is that the inflaton $\phi$ undergoes a super-Planckian 
excursion in field space.
Thereby, it's reasonable to consider the possibility that some extra particle(s) 
with inflaton-dependent masses become light during inflation 
at some specific value(s) $\phi_{\rm pole}$ of the inflaton field.
A simplified model to capture this phenomenon is obtained adding 
a fermion $\Psi$ with a Yukawa coupling $y$ to the inflaton $\phi$,
\beq  \Lag = \frac{(\partial_\mu \phi)^2}{2} - V(\phi)  + \bar\Psi (i \slashed{\partial} - M_\Psi - y \phi) \Psi \eeq
so that the fermion mass is  $\bar M_\Psi=  M_{\Psi} + y \phi$.
In string models, extra gauge vectors $V$ with gauge coupling $g$
can similarly become light at special points in moduli field space,
$\bar M_V^2 = (M_{V} + g \phi)^2$~\cite{Ibanez:1990ju}.
Extra scalars $S$ tend to behave in a different 
way, becoming tachionic after crossing $\bar M_S^2=0$,
giving rise to  `water-fall' inflation.
The masses $M_{S,\Psi,V}$ could be of Planck size.

\medskip

How does the possibility that some extra particle gets massless at $\phi=\phi_{\rm pole}$ 
affect the inflaton action?
Quantum effects due to the light particle
can be computed in QFT by expanding the inflaton field as
$ \phi =\phi_{\rm pole}+ \delta \phi$
around the special value $\phi_{\rm pole}$, obtaining
\beq \Lag_{\rm eff} = \frac{K}{2} (\partial_\mu \delta\phi)^2-
\left[ V_0 +T _0\, \delta \phi + \frac{m_0^2}{2}\delta \phi^2 + A_0\, \delta \phi^3 +
\lambda_0 \, \delta \phi^4 + \cdots\right] +\cdots.\label{eq:Leffpole}
\eeq
If the full theory were renormalizable, the expansion in $\delta \phi$ would stop here and be exact. 
The dominant quantum corrections are log-enhanced and
can be computed by solving the usual RG $\beta$ functions for the
effective couplings $K,V_0, T_0, m_0^2,\ldots$, 
and promoting them into field-dependent couplings 
by setting the RG scale $\mu$ to a value comparable to the light particle mass,
$\mu \sim y\, \delta\phi=y (\phi-\phi_{\rm pole})$, implying
a significant dependence of couplings on $\phi$ around $\phi_{\rm pole}$.
The only relevant parameters around $\delta \phi= 0$ are $K$ and $V_0$,
as the higher order terms in $\delta\phi$ (linear, quadratic, cubic, quartic,...)   vanish at $\delta\phi=0$.

For the moment we focus on the correction to $K$, ignoring the correction to $V_0$.
This could be justified in supersymmetric models, 
where the super-potential does not receive quantum corrections.
The fermionic one-loop correction in $4-2\epsilon$ dimensions and dimensional scale $\mu$ multiplies
the kinetic term of a $\delta\phi$ scalar with momentum $p$ by
\beq 
K \simeq 1+\gamma\left[\frac{1}{\epsilon}+ \ln \frac{\mu^2}{\max( \bar{M}_\Psi^2, p^2)}\right]\label{eq:K1} \eeq
where the coefficient $\gamma$ is known as anomalous dimension of $\delta \phi$.
The anomalous dimension contributes to the gauge-invariant $\beta$ functions of the couplings,
such as $\beta_\lambda = 4\lambda \gamma + \cdots$ for the scalar quartic.
Acceptable large-field inflaton potentials seem to have non-renormalizable form,
making convenient to keep $\phi$ non-canonical, $K\neq 1$.
The scalar one-loop anomalous dimension is $\gamma =(2 y^2-C g^2)/(4\pi)^2$ 
in a theory with a Yukawa coupling $y$ and a gauge coupling $g$.
The coefficient $C$ is gauge-dependent and morally positive,
as the corresponding gauge-independent multiplicative terms in the full $\beta$ functions for the renormalizable couplings are positive.

The renormalisation-group resummation of log-enhanced effects
promotes the one-loop result of eq.\eq{K1} as 
\beq
K_{\rm ren}\simeq 1+ \gamma\ln\frac{\mu^2}{\delta\phi^2}\to 
\exp\left(\int \gamma \, d\ln \mu^2\right)\propto \frac{1}{\delta\phi^{2\gamma}}\eeq
for small constant $\gamma$ and small $\delta\phi$.
So:
\begin{itemize}
\item  A Yukawa coupling induces $K>1$ roughly corresponding to
eq.\eq{Kpole} with a pole with power $p\simeq 2\gamma>0$ i.e.\ dimension $d\simeq 1+\gamma$.
The extra constant $1$ in eq.\eq{Kpole}  accounts for the normal kinetic term, 
away from the special point $\phi_{\rm pole}$ where extra fermions become light.
The parameter $\phi_*$ is  the field range within which the extra particle is lighter than the inflaton,
for order unity couplings $y$.
\item A gauge coupling tends to induce $K<1$ roughly corresponding to eq.\eq{Kpole} with a dip, $p<0$.
 Making the inflaton kinetic term canonical, this is equivalent to add a sudden drop in the inflaton potential.
Samples are plotted in fig.\fig{SampleVSemiPole}.
\end{itemize}
RG equations can be computed in any given particle physics model, 
possibly motivated by other considerations: the fermions $\Psi$ could be 
right-handed neutrinos; the vectors $V$ could be U(1)$_{B-L}$ or SU(5).
The large $\gamma\sim 1$ needed to get significant pole effects during inflation
needs large couplings $g,y \lesssim \sqrt{4\pi}$.
A roughly constant $\gamma$ and thereby a pole-like structure
arises if the RG running towards low energies
enters an asymptotic safety regime.
Banks-Zaks have shown that this can happen in a controllable perturbative regime~\cite{Banks:1981nn}.



\subsection{Extra effects}
Particles that become temporarily light during inflation generate extra effects,
in addition to a quantum correction to the inflaton kinetic function $K(\phi)$.

\medskip

First, quantum corrections to $V(\phi)$ away from $\phi_{\rm pole}$ can damage the flatness of
the inflaton potential.

Second, the vacuum energy $V_0$ in eq.\eq{Leffpole} is large during inflation and,
in a theory with Yukawa couplings $y$ and gauge couplings $g$,
$V_0$ runs at two loops as\footnote{We give order one factors in a concrete example, 
the Standard Model Higgs $H$.
Its potential is usually expanded as
$ V = V_0+M_h^2 |H|^2 + \lambda |H|^4 $
around the special point $H=0$.
Above all SM mass thresholds the only massive parameter is the Higgs mass parameter,
and the SM $\beta$ function for the cosmological constant 
$d V_0/d\ln \mu=  \beta_{V_0}^{(1} + \beta_{V_0}^{(2)} +\cdots$ is
$\beta_{V_0}^{(1)}= 2 M_h^4/(4\pi)^2$ at one loop (which reproduces the well known one-loop potential), 
and
\beq
(4\pi)^4 \beta_{V_0}^{(2)}= (-12 y_t^2+12 g_2^2+\frac{12}{5} g_1^2) M_h^4
+\left[-27 y_t^4+y_t^2(80g_3^2+\frac{45}{2}g_2^2+\frac{17}{2} g_1^2)-\frac{271}{8}g_2^4+\frac{27}{20}g_1^2 g_2^2+\frac{1293}{200} g_1^4\right]V_0
\eeq
at two loops. Some coefficients have been computed in~\cite{1709.02397}.
Notice however that the Higgs $H$ is a special scalar that is accidentally light at inflation end.
The smallness of cosmological constant is a special feature of the SM minimum, that makes
the multiplicative renormalisation of $V_0$ irrelevant.
}

\beq \frac{dV_0}{d\ln \mu} \sim \frac{g^4- y^4}{(4\pi)^4} V_0\eeq
This induces an extra feature in the inflaton potential.

\medskip

Third, in addition to virtual loop effects, particles can be produced during inflation
while their mass $\bar M(\phi)$ becomes
temporarily small as the inflaton evolves in conformal time as $\phi(\tau)$.
Fermion production was studied in~\cite{hep-ph/9910437}:
simple analytic estimates are obtained
assuming that $\bar  M \ll m_0 $ is crossed suddenly,
in a small proper time $\tau \sim k/\bar M'$
where $\bar  M' = d\bar  M/d\tau$.
In such a case, the fermion wave-functions $u_k(\tau)$ with momentum $k$
are significantly distorted from the free-particle solutions $e^{i\omega_k \tau}$
with  $\omega_k^2 =k^2 + a^2 M^2$ up to a maximal momentum
$k_{\rm cr} \lesssim \sqrt{\bar  M'}$ that can be larger than $H$.
Fermions are next inflated away, giving the number density
\beq n(t) \approx \med{\bar\Psi\Psi} = n_* \, \theta(t-t_*) e^{-3H(t-t_*)}\qquad\hbox{where}\qquad
\qquad n_* \sim k_{\rm cr}^3\sim y^{3/2} \dot\phi^{3/2}_* .\eeq
This fermion density back-reacts inducing an extra term in the inflaton equation of motion
$\ddot\phi + 3 H \dot \phi = - V' + y \med{\bar\Psi \Psi}$
that reduces $\dot\phi$~\cite{hep-ph/9910437}.
Unless fermions decays fast, this real effect should be added to the virtual effects.

%

\section{Discussion and conclusions}\label{end}
In conclusion, we have investigated  an inflaton kinetic function $K(\phi)$ featuring a traversable pole. 
This is equivalent to a finite stretching of the inflaton potential, after  transforming the inflaton field into its canonical form.
Section~\ref{sec:formalism} presents the formalism for directly computing inflation in the non-canonical basis.
Assuming a kinetic function of the form of eq.\eq{Kpole}, 
a pole with power $p<2$ is traversable during inflaton, and
can lead to a significantly enhanced power spectrum $P_\zeta(k)$ of
primordial inhomogeneities
at sub-cosmological scales $k$.
Consequently, this can lead to observable scalar-induced gravitational waves and to the formation of primordial black holes. 
Two sub-cases need to be distinguished:
\begin{itemize}
\item If $1\le p < 2$ the canonical potential develops an exact inflection point,
with vanishing first and second derivatives. 
This pole is not traversable in slow-roll approximation.
Nevertheless, a calculation beyond the slow-roll approximation reveals that the pole is classically traversed,
provided that its width coefficient is below a critical threshold.
A big enhancement of the power spectrum arises if the
pole width is tuned to be just below this critical value.
The power spectrum grows as $k^6$, reaches a peak and next decreases.
If $p\gtrsim 1.3$ (depending on the inflaton potential) the decrease is too slow to allow
for a significant enhancement of $P_\zeta$ up to 0.01.

\item If $0<p<1$ the second derivative of the canonical potential diverges at the pole,
that remains traversable even within the slow-roll approximation:
increasing the coefficient of the pole no longer yields an arbitrarily large enhancement of the power spectrum.
Nevertheless, the slow-roll approximation breaks down,
a significant enhancement of $P_\zeta$ still happens for
$p \gtrsim 0.8$ (depending on the inflation potential), 
with a characteristic different shape of $P_\zeta(k)$,
and without significantly tuning the model parameters.

\end{itemize}
Fig.\fig{SamplP} shows example of power spectra.  

In section~\ref{GW} we computed the consequent frequency spectrum $\Omega_{\rm GW}(f)$ of scalar-induced gravitational waves.
As depicted in fig.\fig{PlotGWPoleInflation} these second-order gravitational waves could reproduce the nHz signal
claimed by Pulsar Timing Arrays, or produce signals at higher frequencies.
Fig.\fig{PlotGWPoleInflation} also illustrates the expected astrophysical foreground, that could act as background.

In section~\ref{BH} we computed the consequent mass spectrum $f_{\rm PBH}(M)$
of Primordial Black Holes.
As illustrated in fig.\fig{PlotPBHPoleInflation}, these PBH can
fall  in the asteroid-like mass range where PBH can constitute all of Dark Matter.
Since the mass of PBH has one-to-one correspondence with the peak of the GW signals,
these PBH as DM can be tested via observation of GW peaking in the mHz region, 
to be explored  by LISA.
Alternatively, the sample $P_\zeta(k)$ that reproduces the Pulsar Timing Arrays anomaly
would give heavier PBH, in the testable sub-solar mass range. 
We neglected non-Gaussian corrections, that should be included in more precise computations of the gravitational-wave spectrum and of the PBH abundance.

\medskip

Finally, in section~\ref{theory}, we explored a possible theory
that gives a pole-like structure in the inflaton kinetic function:
quantum corrections originating from Planckian particles with inflaton-dependent mass
(such as fermions with a Yukawa coupling to the inflaton) that become light at some
specific inflaton vacuum expectation value.
In such a case the pole power $p$ is related to the quantum anomalous dimension of the inflaton.
This might give single-field inflation theories with enhanced $P_\zeta$ and mild tuning.
However, as discussed, this theoretical approach tends to introduce additional effects beyond the mere presence of the pole:
dedicated future computations are needed to clarify.
Multiple poles might arise at different values of the inflaton vacuum expectation values,
corresponding to different particles becoming light.
In such a case the power spectrum and the consequent gravitational wave spectrum could exhibit multiple peaks.
If future experiments will advance gravitational wave astronomy, the detection of such signals
could provide insights into Planckian physics.




\paragraph{Acknowledgements}
We thank Chao Chen, Michele Redi and Sotirios Karamitsos for useful discussions.

\end{document}